\documentclass[12pt,preprint]{aastex}

\shorttitle{The Chandra Fornax Survey}
\shortauthors{Scharf, Zurek \& Bureau}

\begin{document}


\title{The Chandra Fornax Survey - I: The Cluster Environment}


\author{C. A. Scharf}
\affil{Columbia Astrophysics Laboratory, Columbia University, MC 5247, 550 West 120th Street, New York, NY10027, USA}
\email{caleb@astro.columbia.edu}

\author{D. R. Zurek}  
\affil{Department of Astrophysics, American Museum
of Natural History, Central Park West at 79th Street, New York, NY
10024} 
\email{dzurek@amnh.org}

\author{M. Bureau \altaffilmark{1}}
\altaffiltext{1}{Hubble Fellow}
\affil{Columbia Astrophysics Laboratory, Columbia
University, MC 5247, 550 West 120th Street, New York, NY10027, USA}
\email{bureau@astro.columbia.edu}



\begin{abstract}

We present the first results of a deep {\sl Chandra} survey of the inner
$\sim 1$ degree of the Fornax cluster of galaxies. Ten 50
ksec pointings were obtained in a mosaic centered on the giant elliptical
galaxy NGC 1399 at the nominal cluster center.
Emission and temperature maps of Fornax are presented, and an
initial study of more than 700 detected X-ray point sources is made. 
 Regions as small as 100pc are resolved. The intra-cluster gas in Fornax
exhibits a highly asymmetric morphology and temperature structure, dominated by
 a 180 kpc extended ``plume'' of low surface brightness, cool ($\leq 1$ keV)
gas to the North-East of NGC 1399 with a sharper edge to the South West.
The elliptical galaxy NGC 1404 also exhibits a cool halo of X-ray gas within
the cluster, with
a highly sharpened leading edge as it presumably falls into the cluster, and a cometary-like
tail. 

We estimate that some $\sim 200-400$ point sources are physically associated with
Fornax. Confirming earlier works, we find that the globular cluster population
in NGC 1399 is highly X-ray active, extending to globulars which may
in fact be intra-cluster systems. We have also found a remarkable 
correlation between the location of giant and dwarf cluster galaxies and
the presence of X-ray counterparts, such that systems inhabiting regions 
of low gas density are more likely to show X-ray activity. Not only does 
this correlate with the asymmetry of the intra-cluster gas but also with
the axis joining the center of Fornax to an infalling group 1 Mpc to
the South-West. We suggest that Fornax may be experiencing an intergalactic ``headwind''
due to motion relative to the surrounding large-scale structure.

\end{abstract}




\section{Introduction}

Low X-ray luminosity galaxy clusters appear to be the dominant
reservoirs of bound matter in the Universe ~\citep{mul00}, and therefore
represent crucial laboratories for studying the state, and history,
of cold and cooling baryons. In the hierarchical structure growth paradigm these
poorer systems also represent the building-blocks of larger, rich
clusters. Ideally we would study them {\it in-situ}, at high redshift, however
this is extremely difficult because of the faintness of their X-ray gas component
and low galaxy density contrast.
We can however study them in great detail at low redshift, and then
by proxie extend much of what we learn about their astrophysics to
earlier epochs.

The relatively shallow gravitational potentials in low-luminosity, or
poor, clusters imply a regime where the energetic influence of
astrophysical phenomenae such as supernovae associated with star
formation and  active-galactic-nucleii (AGN) is comparable to that of
gravity. The observation of an apparent minimum gas entropy in the
core of clusters is one possible consequence of this (~\citet{pon03} and refs therein). However,
the precise interplay of these non-gravitational energy sources with
the intra-cluster medium (ICM), it's evolution, and the member
galaxy population is currently unclear ~\citep{toz01,tor03,voi03,bor04}.

At a distance of about 20.5 Mpc (e.g. ~\citet{fer00}) the Fornax cluster is second only to the Virgo
cluster in the readiness with which high physical resolution multi-wavelength
data may be obtained. With a best resolution of $\approx 1''$, {\sl Chandra}
offers an unprecedented access to X-ray structures as small as 100 pc in
Fornax. In this work we describe the first results from the {\sl Chandra}
Fornax Survey (CFS) which consists of ten mosaiced pointings of 50 ksec
exposure with ACIS-I, covering the innermost $\sim 1^{\circ}$ of the cluster. The aim of
the CFS is to provide an X-ray dataset of sufficently high quality to allow; a
detailed study of the properties of the ICM, its
relationship to the galaxies in Fornax, the X-ray properties of those galaxies
themselves, and the properties of point sources such as X-ray binaries, and accreting black
holes. 

Recent investigations of Fornax have revealed several characteristics
of this cluster which point towards a highly complex
environment. These include; a possible intra-cluster stellar
population amounting to at least 10-40\% of the total cluster stellar
mass ~\citep{the97,cia02,kar03,nei04}. Secondly, a population of ultra-compact
intracluster stellar systems ~\citep{dri03} often referred to as
ultra-compact-objects or dwarfs (UCOs or UCDs) which are not
readily classifiable as either galaxies or globular clusters. Thirdly,
approximately 1 Mpc to the south-west of the cluster center is an
infalling poor group, suggesting significant dynamical activity in
Fornax ~\citep{dri01}.

In this paper we present the initial results from the CFS. In \S 2 we describe
the basic data processing, in \S2.1 we present multi-band images of the CFS,
and in \S2.2 we present an estimated gas temperature map of Fornax together
with confidence regions and an example of narrow-band imaging designed to
indicate relative metal abundances in the ICM. In \S3.1 we discuss the
infalling galaxy NGC 1404, and in \S 3.2 the tentative motion of NGC 1387 and
the possible truncation of its hot IGM by the surrounding Fornax ICM. In \S 4
we present first results from efforts to compile an X-ray source catalog of
over 700 detections in the CFS, together with X-ray color-color
classifications. In \S 4.1 we discuss the X-ray counterparts to known bright
Fornax galaxy members, and in \S 4.2 we present X-ray measurements of
counterparts to two possible intra-cluster globular clusters in the outermost
regions of the central galaxy NGC 1399. In \S 5 we present an analysis of what
is likely a distant, hot, background cluster of galaxies seen to the immediate
East of NGC 1399. Finally, in \S 6 we summarize the results presented here and
discuss some of the implications for the overall Fornax environment.

\section{Data and Analysis}

The data for the {\sl Chandra} Fornax Survey were obtained
sequentially from the period of May 20 2003 through to June 6 2003.
Sequence numbers were 800324 to 800333. Due
to increased solar activity the observations were halted within this
period, and sequence 800331 was interrupted and restarted once the
solar activity had died down. Exposure times ranged between 45 and 48
ksec.  A schematic of the mosaic strategy is given in Figure 1,
overlaid on an optical {\sl Digitized Sky Survey} image of Fornax. The mosaic was
designed to provide a) contiguous coverage over the cluster core and b) high resolution
(on-axis) data for the brightest galaxies in the Fornax core. Nine fields provided full
coverage and a tenth was positioned to double the exposure in the region
between NGC 1399 and NGC 1404, and ensure that the leading edge of NGC 1404 (\S 3.1) was
not at a chip gap.

 All data were taken using the VFAINT mode, and were reprocessed to
take advantage of the increased background discrimination this
offers. The Charge Transfer Inefficiency (CTI) correction was also
reapplied. After filtering out periods of high background, the range
of usable exposure was 40.8-46.6 ksec across all fields. Gain
corrections were also applied to the data. In addition to  the ACIS-I
0,1,2, and 3 chips, the ACIS-S S2 (chip 6) was switched on, although
we do not consider those data here.

\subsection{Images}

Soft (0.3-1.5 keV), medium (1.5-2.5 keV), and hard band (2.5-8 keV) images of
the full mosaic are presented in Figures 2,3, and 4. Exposure maps were
constructed in these bands for each individual ACIS-I field using a spatial
binning factor of 4 ($2''$) and spectral weights calculated assuming a thermal
MEKAL spectrum of 1.2 keV and 0.3 solar metal abundance. Each field/band was
binned by a factor 4 and then smoothed using the adaptive smoothing algorithm {\sl
CSMOOTH} limited to a maximum smoothing scale of $10''$ . The smoothing scales 
generated for each image were then applied to
the band appropriate exposure maps, all smoothed images and 
 exposure maps were then merged, and an exposure-corrected mosaiced image was
produced in each energy band. It should be noted that some residual structures
associated with the field overlap regions are still apparent in the soft band
image, and to a much lesser degree in the medium and hard images. This is in
part due to the imprecise nature of the exposure map spectral weighting, which
necessarily assumes a uniform spectral model, and in part due to the inherent
difficulty of merging two independently smoothed and exposure corrected
versions of the same part of the sky, containing low surface brightness emission - 
as necessitated by current software
limitations. We will address these issues in subsequent work on the CFS image
maps. 

Several features are notable in Figures 2,3, and 4. First, the soft band image
clearly shows a remarkably asymmetric morphology of the low density Fornax
ICM, which appears to be swept to the North-East. Within the higher density
region surrounding the central giant elliptical NGC 1399 there is a further
asymmetry such that NGC 1399 appears offset to the North-East relative to the
ICM core, which extends further to the South-West. There is also clearly
complex emission in the brightest regions of NGC 1399, probably associated
with the radio-jet structure in this galaxy (Paolillo et al. 2002).  Second,
the galaxy NGC 1404 exhibits a spectacular ``cometary'' X-ray morphology, with
a sharp North-West edge and a wisp-like emission tail
extending some $5'$ (30 kpc) to the South-East (see \S3.1 below). Finally, in
addition to the point source populations which are seen in all three bands
there is an extended source approximately $8.7'$ (52 kpc) to the East of NGC
1399 which is present at all energies. We do not currently know the precise origin of
this emission, although it appears very likely that it is a hot, distant,
background galaxy cluster (see \S5 below).

\subsection{Temperature Mapping}

The gas temperature map shown in Figure 5 was produced using a technique with
similarities to the multiple-band fitting approach of ~\citet{mar00} and
~\citet{vik01}. A grid of {\sl MEKAL} gas spectral models is generated varying the
gas temperature $kT$ from 0.1 to 12 keV in steps of 0.1 keV with abundances
set at 0.3 solar and redshift set to zero, and a Galactic foreground absorbing column
of $nH=1.4\times 10^{20}$cm$^{-2}$. The models are generated using {\sl XSPEC} and
an on-axis {\sl Chandra} response from 0.3-7 keV (variations in response with
off-axis angle have negligible effect on the results presented here). A global
background spectral model (generated using the on-line ACIS-I background
files) is added to each grid model and the result is normalized to a single
integrated count-rate.  Bright sources are removed from the CFS data and
photons are then spatially binned to $32''$ pixels; counts per pixel range
from $\sim 40$ to over 2000 in the 0.3-7 keV band. Within a given pixel each
photon is assigned a unique probability for being drawn from a spectral model.
In other words the normalized spectral model is treated as a probability distribution.
The cumulative probability ($P$) that the photons in a pixel are drawn from a
given model ($kT$) is then calculated and the maximum of this $P(kT)$ is
found, thereby yielding the ``best fit'' $kT$. The resultant temperature map
is then spatially smoothed by a Gaussian of width $16''$ to wash out low level
noise features. A full discussion of this technique, and examples of its
calibration against conventional fitting approaches will be presented by
Scharf (2004, in preparation). The method was originally designed in
combination with a spatial source detection algorithm to act as a
computationally efficient optimal matched-filter for detecting low density,
low temperature gas, and it is therefore quite robust when applied to datasets
with good photon statistics, such as the Fornax data presented here. 

The estimated temperatures in the vicinity of the central galaxy NGC 1399 are
in excellent global agreement with earlier measurements from ROSAT ($1.3 \pm
0.05$ keV, ~\citet{jon97}, $1.1 \pm 0.1$ keV, ~\citet{ran95}), ASCA ($1.2\pm
0.04$ keV, ~\citet{fuk96}) and more recently with XMM-Newton (0.9 \& 1.5 keV
components, ~\citet{buo02}). As a guide to the robustness of the temperatures
estimated here, Figure 6a shows overlaid contours of total (non
exposure-corrected) photon counts. Regions outside the lowermost contour
should be considered as noise and/or background dominated (reflected in
the systematically higher estimated $kT$'s).

In Figure 6b we also present an example of a narrow band X-ray image.  Photons
were extracted in a 140 eV wide band straddling the Fe-L complex (600-740 eV)
and an off-band (460-500 eV). The off-band data were gaussian smoothed with a
kernel radius of $10''$ and an un-sharp masked image was created by
subtracting this smooth off-band data from the on-band image and smoothing the
residual map with a $1''$ kernel (c.f. broad band unsharp masking, e.g. ~\citet{fab03}).
 The resulting image represents enhancements in flux in the Fe-L complex
band relative to the off-band and is therefore sensitive to a combination of
higher Fe abundances and lower $kT$ (which increases the relative Fe-L
emission in these $\sim 1$ keV temperature ranges). The halo of NGC 1404 is
the most significant feature in this image, suggesting that it may have a
higher Fe abundance than the surrounding ICM. More detailed spectral modeling
will be presented in a subsequent work.

\section{Galaxy Motions}

\subsection{NGC 1404}
 
Previous studies of the X-ray morphology of the elliptical galaxy NGC 1404
have indicated the presence of distortion/elongation in the emission along an
axis towards NGC 1399, strongly suggesting infall ~\citep{jon97,pao02}.
Dynamical studies of Fornax have also suggested that NGC 1404 may be part of a
clump (less significant than what is commonly considered major substructure) projected onto the Fornax core
and infalling with a line-of-sight velocity relative to NGC 1399 of $450$ km
s$^{-1}$ ~\citep{dri01}. In Figures 2 and 5 there is excellent support for
this picture. NGC 1404 clearly has a very sharp NW edge in emission, combined
with a ``tail'' of soft emission trailing to the SE. Furthermore, the
temperature map reveals a dramatic plume of cooler material to the immediate
SE.  In Figure 8 we plot the soft band surface brightness profile of NGC 1404
in a pie-wedge set of circular annuli towards the NW (Figure 7). The inner
$40''$ (4 kpc) of emission is very well fit by a single $\beta$ model emission
distribution, which continues to be a good fit to approximately $60''$ (6
kpc). The best fit $\beta$ (0.5) and core radius ($5.1''$) are in excellent
agreement with those obtained from a fully azimuthally averaged profile using
ROSAT HRI data ~\citep{pao02}. Beyond 6 kpc, however, the profile drops rapidly,
by more than a factor $ 10$ in 2 kpc. In addition (Figure 5) the gas
temperature rises from $\approx 0.6$ keV to $\approx 1.6$ keV. In these respects the
situation appears very similar to that of the larger scale ``cold-fronts''
observed in massive clusters ~\citep{vik01,mar00}.  

We now apply a crude
analysis to understand the implications of this structure, a more detailed
investigation will be presented elsewhere. Following ~\cite{vik01} the ratio
of pressure in the free stream (external ICM) to that at the stagnation
point at the leading edge of a moving cold cloud is a function of the
adiabtaic index and Mach number of the free stream. From a deprojection of the
best fit $\beta$ model the approximate gas density at the $70''$ (7 kpc) cold
edge is $n_e\approx 4\times 10^{-3}$cm$^{-3}$ versus $n_e\approx 0.6 \times
10^{-3}$cm$^{-3}$ in the ambient medium. This implies a pressure ratio between
the NGC 1404 halo and the ICM of $p_o/p_1\simeq 2.5\pm 0.5$, corresponding to
a Mach number of $M_1\simeq 1.3\pm0.3$. NGC 1404 is therefore moving at or
slightly higher than the ICM sound speed. The estimated velocity is $\simeq
660\pm 260$ km s$^{-1}$.

 If NGC 1404 is indeed moving supersonically then a
bow shock is expected to form at some distance ahead of the motion. Because of
the projected proximity to the massive galaxy NGC 1399, its X-ray halo, and the core of
the cluster potential in the direction of motion, it is likely that any such
bow shock will be hard to detect directly. Indeed, no such feature is readily
apparent. Using the Rankine-Hugoniot jump conditions and the estimated Mach
number, we can however estimate the likely temperature jump as being a factor
$\sim 1.2$. At close to sonic velocities we would also expect compression of
the ambient ICM in the stagnation region immediately ahead of the cooler halo
of NGC 1404, and this may correspond to the slight flattening of the profile
visible in Figure 8 at $55-70''$ (5.5-7 kpc). 

We note also that NGC 1404 may be producing a Bondi-Hoyle (B-H) gravitational
wake - which could mimic aspects of a ram-pressure stripped tail, while
actually being the gravitationally focussed ICM ~\citep{ste99,sak00}.  It is
also possible that some combination exists, such that the halo material of NGC
1404 is mixing with the ICM wake material. Visible B-H wakes are most likely
for subsonic motion of a galaxy through a relatively cool ICM, and in the case of
NGC 1404 at least one of these criteria is met in as much as the ambient ICM
is quite cool ($\sim 1$ keV). ~\citet{sak00} performed a simulation of NGC 1404
and its possible B-H wake and determined that with a velocity in the plane of
the sky of 300 km s$^{-1}$ the B-H wake could extend to 17 kpc;  with a 500
km s$^{-1}$ motion it could extend to 20 kpc. From the CFS data we estimate the
length of the NGC 1404 tail to be (beyond an azimuthally averaged mean radius
of $\approx 100''$ (10 kpc)) at least 30 kpc. As ~\citet{sak00} notes, a wake
exceeding 20 kpc could be due to a greater infall velocity, and our above estimate
of $\approx 700$ km s$^{-1}$ is certainly consistent with this. 

Earlier dynamical and X-ray studies of NGC 1404 have suggested that it is
likely undergoing its first ``fall-through'' in the Fornax system. If the
emission tail seen here in the CFS is indeed ram-pressure stripped material
then this picture is clearly supported. From these X-ray results we now also
have an estimate of the velocity relative to the ICM/cluster frame close to
the plane of the sky. This is comparable to the purely radial optical redshift
velocity of 450 km s$^{-1}$ ~\citep{dri01}. Combined, these velocity vectors
indicate an absolute upper limit for motion towards the cluster core of $\approx
800$ km s$^{-1}$. Further deprojection modeling of the NGC 1404 emission may
allow us to better constrain the velocity vector angle relative to the line of
sight, and ultimately constrain the orbital parameters for this galaxy in
the cluster potential.

\subsection{NGC 1387}

As the third brightest optical galaxy after NGC 1399 and NGC 1404, the galaxy
NGC 1387, approximately 120 kpc to the West of NGC 1399, also exhibits a 
significant X-ray halo. The optical classification of
NGC 1387 is slightly complex. There is an elongation of the bulge within an
extended, nearly circular, very faint outer envelope, resulting in an SB
classification. However, the envelope is definite and pronounced,
characteristic of an S0 ~\citep{san94}. 

Owing to the design constraints of the CFS mosaic the emission from this
galaxy falls across one of the ACIS-I chip gaps and so its interpretation is
complicated by the need to carefully incorporate exposure corrections.
There are no
signficant morphological features in the
soft emission map of Figure 2 and in the raw photon distribution.
However, in the temperature map of Figure 5,
there is evidence for a cool ``tail'' of gas similar to, but much weaker than,
that seen in NGC 1404.  This feature extends to the immediate NE of the
galaxy, suggesting motion in the plane of the sky towards the SW. In Figure 9
the NE and SW emission profiles are plotted for NGC 1387 using pie-wedge circular
annuli (see Figure caption). There is no significant difference between these
profiles; both show evidence for a point source contribution within 1 kpc and
are reasonably well fit by $\beta$ profiles between 1  and $\approx 10$ kpc.
Both profiles do however show evidence for truncation beyond $\approx 14$ kpc,
dropping by a factor $\approx 6$ over only $\approx 2-3$ kpc.  Similar confinement
of cooler media by an external ICM has been seen in the Coma cluster by
Vikhlinin et el. (2001), and the possible adiabatic compression of a galaxy
gas halo by an external ICM has been further considered by ~\citet{fin04}. 

The absence of any notable emission morphology akin to that of NGC
1404 suggests that any motion of NGC 1387 is likely subsonic, combined
with a lower ICM density at its location. The presence of a low-temperature
tail may therefore point towards a Bondi-Hoyle accretion wake, although the
lack of an obvious enhancement of gas density may argue against this.

\section{The source catalog}

To produce an initial source catalog, we merged the CFS data with the
existing archival {\sl Chandra} data of NGC 1399. Specifically, we merged the ACIS-S
S3 chip data from the {\sl Chandra} archive with the full CFS dataset.  The combined
exposure on NGC 1399 is approximately 145 ksec (two ACIS-I pointings, plus 55
ksec from ACIS-S). We ran the WAVDETECT source detection algorithm on the
0.3-8 keV combined image at two different binning scales ($1''$, and
$2''$ pixels) using the default parameters. The use of two image binning
scales allows the algorithm to detect both the smallest angular scale sources
and those of larger extent and/or lower surface brightness. We then merged the
resulting lists visually, removing duplicates and selecting the resolution of
the source detection which best enclosed the source flux. A total of 771
sources were thus obtained (Figure 10). 

We then extracted source counts in three bands: soft (0.3-1.0 keV), medium
(1.0-2.0 keV), and hard (2.0-8 keV) using the broad band source apertures.
Note that these energy bands are slightly different from the bands used in
Figures 2---4, and are chosen to match those used by ~\citet{sor03} and
~\citet{pre03} for source classification. Owing to the variation in
response between the ACIS-I and ACIS-S instruments, we here present results of
source counts/colors from the CFS ACIS-I data only, although the set of source
apertures is that derived from the combined CFS plus archival dataset. A total
of 437 sources were obtained in the CFS with non-zero counts in all three
energy bands.  We plot the color-color diagram for this subset in Figure 11,
following the color definitions of ~\citet{pre03} and ~\citet{sor03}. 
Two colors are defined, a ``hard'' color: $(H-M)/(H+M+S)$ and a ``soft''
color: $(M-S)/(H+M+S)$, where $S,M,$ and $H$ are the soft, medium, and hard
total counts respectively. 
Typical error bars on the colors are large, of the order $\pm 0.1-0.4$ color
units, owing to the relatively low photon count of the majority of sources.

In Figure 11 we also plot several curves corresponding to different
spectral models. The curves trace a function of neutral column absorption,
from the known foreground Galactic column of $nH=1.4\times 10^{2}$ cm$^{-2}$ to
an arbitrary value of $2\times 10^{22}$ cm$^{-2}$. Three power law
spectra are plotted with photon indices of $\Gamma=0$, 1.4, and 2.0. Many
of the background AGN sources (see below) are likely to correspond to
the range of colors covered by these power law spectra, as well as X-ray binaries (XRBs)
in the hard state ($\Gamma=1.3-1.7$).
 We also plot curves for: a $kT=0.1$ keV black-body, typical of a supersoft
source, a $0.5$ keV disk black-body, corresponding to an XRB in its quiet state,
and a $0.5$ keV thermal spectrum akin to a supernova remnant/supernova driven
outflow source or other soft emission. The 437 sources with non-zero counts in
all bands clearly cover a wide range of potential spectral types

A significant, possibly majority, fraction of point sources in the CFS are
expected to be background AGN. We estimate the {\em maximum} number count of
such sources from the extra-galactic LogN-LogS curves as determined from the
{\sl Chandra} deep fields ~\citep{ros02}. In the {\em absence} of the
Fornax ICM emission the approximate flux limit for the non-overlapping CFS
exposures is $\simeq 10^{-15}$ erg s$^{-1}$ cm$^{-2}$ for a $\sim 3\sigma$
detection of a power law point source on-axis (where the detection
significance is given by $s/\sqrt{s+n}$ in the usual Poisson limit). Given a
CFS size of $\sim 0.7$ deg$^{2}$ we then estimate a mean background source
count of $\sim 560$ (0.3-8 keV). However, this assumes no ICM contribution to
the effective background above which sources must be detected. We have
estimated a conservative mean for the ICM contribution to the 0.3-8 keV
background of approximately a factor 2 times the normal quiescent ``blank
field'' background for ACIS-I. For a typical source detection aperture of
$\simeq 4''$ radius this implies a net increase in the source flux detection
limit (for a $3\sigma$ detection assuming Poisson errors) of approximately
30\%. Owing to the steepness of the extragalactic $\log N-\log S$ function this translates to a new mean
background count of $\sim 420$. These numbers should be treated with caution,
and the additional depth of the CFS in overlap regions (plus the inclusion of
the archival NGC 1399 data) will boost the observed background count rates
above these estimates (although surrounding NGC 1399 the ICM background is as
much as two orders of magnitude larger than considered here). In summary, it
is likely that some $210-360$ of the detected sources are physically
associated with the cluster.

\subsection{X-ray/optical galaxy counterparts and the morphology of Fornax}

We have compared the distribution of X-ray sources in the CFS with a catalog
of spectroscopically confirmed Fornax member galaxies from ~\citet{kar03}.
 The optical catalog consists of 92 galaxies spectroscopically confirmed as
cluster members over a $4\times 3$
deg$^{2}$ region down to a limiting magnitude of $b_J\sim 19.8$. Of these galaxies 26 are within
the CFS region. Figure 12 shows the positions of all X-ray sources overlaid on
the soft-band CFS image, and the locations of the cluster member galaxies.
We searched within a 3 kpc ($30''$) radius of each galaxy center for X-ray
counterparts, and such coincidences are labeled by arrows in Figure 12.  The
distribution of potential counterparts is strikingly asymmetric. 
There are 11 galaxies (excluding NGC 1399, but including NGC 1404 and NGC
1387) with possible X-ray counterparts in total (42 \% of all cluster galaxies within the CFS), and 9
of these lie to the west/south-west of NGC 1399, and 1 is to the far north-east.  The range of optical
morphologies of the galaxies with potential X-ray counterparts are 4 S0 or
SB0's, 5 dS0 or dE's, and 3 E's (including NGC 1399 and 1404), following the
classifications of ~\citet{kar03} and ~\citet{dri01}. The X-ray
colors of the counterparts indicate that the sources range from soft to
hard spectra and may therefore include both power law (e.g. XRB's, AGN) and
thermal (e.g. starburst, gas halo) spectral types. 

Although the numbers are small, the distribution of X-ray active
galaxies in Fornax suggests an anticorrelation with the distribution
of the ICM.  The obvious implication is that the galaxies are strongly
influenced by interaction with the ambient medium, and perhaps that
these galaxies belong to a different population, such as field objects
now entering the cluster environment. The idea that such an influence
exists is in itself not new,  but the majority of studies on the
influence of the cluster/density environment on member galaxies have focussed
on galaxy properties such as HI content (e.g. ~\citet{sol01}) and optical stellar
activity (e.g. ~\citet{bal04}). In more recent works however, ~\citet{pog04}
and ~\citet{fin04} have investigated respectively the spatial
distribution of starburst/post-starburst galaxies and that of X-ray
luminous galaxies in the Coma cluster.

~\citet{fin04} conclude that the X-ray luminosities of galaxies in
Coma is quenched by a factor of almost 6 compared to those of galaxies in the
field, and that this must be linked to a reduction in star formation activity
of galaxies in cluster environments. By comparison, ~\citet{pog04}
find that the youngest and strongest post-starburst galaxies (E+a spectral
types) are located close to the edges of infalling substructures, suggesting
that interaction with the ICM is responsible both for a previous starburst
episode and for subsequently quenching this activity. The implications for our
present study of Fornax are clear - the X-ray active galaxies are likely in
the process of being strongly influenced by the ambient medium. They are
however not yet ``quenched'', and may indeed be stimulated by infall through
the less dense West/South-West Fornax ICM. 

X-ray studies of Fornax over a wider field have identified a
significant sub-cluster (including NGC 1316), likely infalling and approximately
1 Mpc to the South-West of NGC 1399 ~\citep{dri01}. The
approximate location and size of this system is shown in Figure 13.
 The geometry
suggests that the Fornax core and the sub-cluster may lie along part of a
filamentary structure which is collapsing and flowing in towards the common
center of mass. In this picture, the morphology of the ICM seen in the CFS is
consistent with either the agglomeration of IGM from the North-East and/or the
presence of a relatively gentle ram-pressure tail as the NGC 1399 potential well
drifts along the filament to the South-West, towards the common center of
mass. In either scenario, the galaxies between NGC 1399 and the infalling
sub-cluster are uniquely located in a region bounding two significant
structures and containing a moderate density 1-2 keV ICM (a factor $\approx 10$
less dense than the North-Eastern regions).

\subsection{X-ray counterparts to outlying NGC 1399 globular clusters}

We have made an initial investigation of possible X-ray counterparts to a
catalog of compact objects in the Fornax core. This catalog consists of 5
ultra-compact objects (UCO's, ~\citet{hil99}) and 18 objects from a
spectroscopic survey of bright compact objects extending into the outer
regions of NGC 1399 ~\citep{mie02}. We have found 3
unambiguous matches with our X-ray source catalog, with X-ray/optical offsets
of $\leq 0.5''$. Following the nomenclature of ~\citet{mie02}
 these objects are FCC 208 (dwarf elliptical, dE, N), FCOS 1-060
(Globular cluster), and FCOS 2-073 (Globular cluster). None of the UCO
candidates were detected as X-ray sources. The location of the two globular
cluster (GC) candidates is illustrated in Figure 14. Both are significantly
distant from the center of NGC 1399, at 60 and 79 kpc for FCOS 1-060 and FCOS
2-073 respectively. 

We have extracted source counts from $10''$ diameter apertures around each GC,
FCOS 1-060 lies in a region of double exposure in the CFS (90 ksec) while FCOS
2-073 has a 45 ksec exposure.  The background subtracted source counts and
rates for FCOS 1-060 and FCOS 2-073 use a mean background estimated from 10
apertures randomly placed within 1 arcmin of these objects, avoiding other
sources. The X-ray properties are given in Table 1. The color indices have
errors of $\sim \pm 0.3$. Based on the color-color plane (Figure 10) we then
tentatively assign spectral types: FCOS 1-060, Disk Blackbody, $kT\leq 0.5$
keV, and FCOS 2-073, Power law, Galactic absorption, photon index 1.4. The
intrinsic luminosities (Table 1) of these two objects are then estimated based
on these spectral models.

The observed X-ray luminosities are consistent with those measured in the
inner GC population of NGC 1399 ~\citep{ang01}.  It is intriguing that both
objects have X-ray luminosities consistent with the Eddington limit luminosity
for spherical accretion onto a 1.4 $M_{\odot}$ neutron star ($2\times 10^{38}$
erg s$^{-1}$), which suggests that it is conceivable that we are detecting
just {\em one} low mass X-ray binary (LMXRB) system in each GC. However, as
~\citet{ang01} point out, in our galaxy and in 500 GC systems in M31 there are
{\em no} GC's with X-ray luminosities exceeding $1\times 10^{38}$ erg
s$^{-1}$. Within our Galaxy all LMXRB's have luminosities between $\sim
10^{36}$ and $\sim 7\times 10^{37}$ erg s$^{-1}$. Thus it might reasonable to
assume that a population of anything between 2-200 LMXRB's per GC would also
produce the observed luminosity. 

Such a large number of LMXRB's in a GC indicates that the efficiency of producing
such systems is greatly enhanced compared to the GCs of the Milky Way or
M31. The stellar dynamics of the GCs in question must therefore be different than
that of the typical GC in the Milky Way (or M31). It has been shown from
N-Body simulations ~\citep{gia85} that the number of stars initially in binaries
greatly enhances stellar interactions and the tightening of binary orbits.
This process eventually, after many interactions, produces mass exchange
systems such as LMXRB's. Large N-Body simulations of these systems($>$ 100,000 stars) have
only just begun to produce results, however, these simulations do {\em not} appear to produce
hundreds of LMXRB's. \footnote{Prior to the existence of GRAPE-6 computers ~\citep{mak03},
simulations this large were rarely undertaken.}

The question was raised by ~\citet{ang01} that the dynamics of the GCs
in NGC 1399 would therefore have to be different to produce so many LMXRB's. While it is
hard to imagine a scenario where the dynamics of a GC would be so different
from that of the Milky Way or M31, it should be stressed that the Fornax
system is a cluster of galaxies that has undergone many galaxy-galaxy
interactions and mergers. As well as an increase in the specific frequency
of GCs due to the merger of populations, GC formation may also occur
during the merger of giant galaxies ~\citep{ash92,zep93}. If the
dynamics were different for the GCs existing prior to mergers in Fornax the GCs
must have been tidally shocked during the galaxy mergers - stripping a large
fraction of the low mass stars from a GC and causing it to undergo core
collapse and creating many new tight binary systems ~\citep{kun95,gne99a,gne99b}. If, however, the
systems we are seeing are the GCs created {\em during} the galaxy mergers, then the
number of short period binaries must have been greatly enhanced from that of
GCs formed prior to the mergers. It would then also be likely that they would 
be younger than the general GC population. We will pursue this question in a future study.

Another possibility is that a fraction of all GCs contain black holes
(BHs) of intermediate mass (100 - 10,000 M$_{\odot}$). The evidence for BHs in GCs
of the Milky Way and M31 ~\citep{geb02,ger02,col03} is weak at this time ~\citep{mcn03,bau03},
though suggestive. If
a fraction of all GCs have intermediate mass BHs then at any given time
a fraction of that number must be in a high state due to the disruption of
stellar material. The total number of GCs that may have a BH in
them is about 1 to 5 in both the Milky Way and M31, which is $\sim$1\% of the
total GC population. NGC 1399 has more than 6000 GCs and if the fraction
is the same that would suggest that about 60 GCs in NGC 1399 would have BHs. Of
the possible GC BH candidates in the Milky Way, or M31, we know that none are
X-ray bright and therefore it is likely that less than 10\% of the BHs in Fornax GCs
would be active now. That suggests that about 6 GC systems with a BH would
be X-Ray bright now in NGC 1399. This is far fewer than the number found
by ~\citet{ang01}, and suggested here in the CFS. If low-intermediate mass BHs
are indeed the source of the observed high X-ray luminosities then the number of Fornax
GCs with BHs must be enhanced over that of the Milky Way and M31, which presents
similar difficulties as invoking an enhanced production of LMXRBs.

Future studies will look into population differences of the GC systems and
determine if any correlation can be found with optical properties and X-ray
properties.

\section{A distant background cluster of galaxies ?}

As discussed in \S 2 above, we have found an extended X-ray source
approximately $8.7'$ (52 kpc) to the East of NGC 1399, at a nominal position of
(03 39 12, -35 26 09) (J2000) with a soft band extent of $\approx 1.4'$ diameter.
This source is clearly seen in all three energy bands, suggesting that it is
spectrally much harder than the Fornax ICM. In Figure 15a the soft band
(0.3-1.5 keV) raw image (binned to $4''$ pixels) is shown with contours of the
adaptively smoothed, exposure corrected, soft band image (Figure 2) overlaid.
In Figure 15b a deep R-band image is shown centered on this position. This
data was taken using the Mosaic camera on the CTIO 4m during December 1999
~\citep{nei04} and has a magnitude limit of approximately $R\sim 23$. 

The moderately bright star towards the image center is unlikely to be
associated with this diffuse emission. There is a clear association of faint
objects ($m_{R}\sim 22-23$) directly coincident with the extended X-ray
emission, including what could be two cD galaxies. 

Analysis of the source spectrum is somewhat complicated by the source position 
in the overlap region of two pointings, but the combined exposure time is 
approximately 90 ksec. We have extracted the source photons in a circular
aperture of 2 arcmin diameter  from both pointings. We have generated
ACIS-I responses for both pointing and then formed a weighted combined
response, and co-added the source photon lists. The total, background
subtracted source photon count is approximately 700 in total (count rate
$7.8\times 10^{-3}$) in the 0.3-10 keV band. The background spectrum is
obtained from regions immediately surrounding this source and therefore
contains the intracluster emission of Fornax.
Limiting the energy range to
0.3-7 keV we have used {\sl SHERPA} to examine possible spectral models for
the emission. Fixing the foreground absorption to the Galactic value of $nH=1.4\times 10^{20}$ cm$^{-2}$
we fit a thermal {\sl MEKAL} model with fixed abundance of 0.3 solar and
free temperature, redshift and normalization using the spectrum grouped
to a minimum of 20 counts per bin and a Chi-Gehrels statistic. The best fit
model yields $kT\simeq 25$ keV, $z=0.6$ and a reduced Chi-Gehrels statistic
of $0.7$. Alternatively, the data can be fit with a slightly poorer
reduced statistic by a single power-law model, with Galactic foreground
absorption and a photon index of $\Gamma=1.3$. A power law model
including a redshifted absorption component as well as Galactic
absorption yields a better statistic but for $\Gamma=1.9$ and
an absorber of $\sim 6\times 10^{22}$ cm$^{-2}$ at $z=3$, which seems
physically unlikely for extended emission.

We therefore tentatively identify this object as CXOU J033912.0-352609, a likely 
hot, massive cluster of galaxies at $z>0.3$ (based also on optical magnitudes), although
the true gas temperature is probably significantly lower than the 25 keV fit here.

\section{Summary and Discussion}

We have described the aquisition and initial analysis of the  most detailed wide-field 
X-ray data on the Fornax cluster to date. 
In presenting some of the initial results of the CFS here we have attempted
to illustrate several key aspects of the Fornax cluster environment which are
now accessible for further investigation. Namely: (1) there is clear evidence of
interaction between at least 2 galaxies (NGC 1404 and NGC 1387) and the
Fornax ICM. In the case of NGC 1404 we have obtained the first quantitative
constraint on it's motion perpendicular to our line-of-sight, opening the
way to constraining it's orbital configuration when combined with existing
redshift information. (2) the Fornax ICM has a clearly asymmetric
morphology which we suggest may be related to the larger scale dynamics 
of this region, in which we are perhaps witnessing the coalescence of Fornax
with an infalling group along a 1 Mpc filamentary structure. (3)
possibly related to this ongoing growth, and almost certainly related to
the local ICM density environment we find that the majority of X-ray 
active Fornax galaxies are distributed away from the bulk of the ICM, and
between the Fornax core and the likely infalling structure. (4) we detect 2 
globular cluster candidates which may be part of the outer structure of NGC 
1399, or may be intracluster systems (formerly part of this, or another galaxy).
Their X-ray luminosity suggests either a population of low-mass X-ray binaries or
possibly an intermediate mass black hole.

The morphology of the Fornax ICM emission bears some resemblance to the
phenomena of cold-fronts seen in other, massive, clusters on scales of
some 0.5 Mpc (e.g. ~\citet{vik01}). However, in these situations the
density of the gas environment is significantly higher than that seen
in Fornax, and much more easily attributable to cool, dense, substructure
passing through a warmer, dense ICM. Furthermore, the proto-typical
cold fronts of Abell 2142 ~\citep{mar00} and Abell 3667 ~\citep{vik01}
are not physically close to the central cD galaxy of those clusters, whereas
in Fornax NGC 1399 appears to be close to the ``head'' of the cometary
type emission pattern. The striking temperature structure of Fornax
also shows evidence for a much more extended cool component than
is perhaps present in these richer systems.
There is in addition an apparent offset between the location
of NGC 1399 and the central core of ICM emission, where NGC 1399 is shifted some
2 arcmin to the North-East relative to the apparent geometric center of the
core plateau. More detailed investigation of the gas density and temperature
distribution is required in this region, but it seems plausible that 
there is relative motion between NGC 1399 and this region of emission.

As discussed in \S4.1 it appears likely that the X-ray morphology of
Fornax as seen in the CFS may be attributed to the ongoing coalescence
of structure. In particular, the infall of the group identified by
~\citet{dri01} suggests that the Fornax core may be experiencing an
intergalactic ``wind'' due to its relative motion with respect to
structure along this infall axis. Unlike the case of strong cold
fronts in the denser environment of massive clusters, this may
represent a gentler phenomenon, associated directly with the intergalactic
medium in the surrounding large scale structure.

Finally, as part of the CFS project, processed maps and source
catalogs will be made available and continously updated via the
Data Collections of the Department of Astrophysics at the
American Museum of Natural History. \footnote{http://research.amnh.org/astrophysics/collections.html}

\acknowledgments

We thank D. Helfand and F. Paerels for useful discussions, E. Gotthelf
for invaluable help in constructing the image maps, D. Neill for
providing  deep optical images, and M. Markevitch of the CfA/CXC for
his expert assistance in the observational preparation.
C.A.S. acknowledges the support of NASA/{\sl Chandra} grant SAO
G03-4158A and  the Columbia Astrophysics Laboratory. D.R.Z. acknowledges the
support of NASA/{\sl Chandra} grant SAO G03-4158B and the American
Museum of Natural History.
Support for this
work (M. B.) was provided by NASA through Hubble Fellowship grant
HST-HF-01136.01 awarded by the Space Telescope Science Institute,
which is operated by the Association of Universities for Research in
Astronomy, Inc., for NASA, under contract NAS~5-26555.

\clearpage

\begin{figure}
%\plotone{f1.eps}
\caption{Schematic of the Chandra ACIS-I mosaic strategy. 10 ACIS-I fields (each consisting of 4 CCDs) are 
overlaid on a DSS image of the Fornax cluster, centered on NGC 1399 with NGC 1404 to the
immediate South-East. Boxed circle symbols indicate the nominal on-axis position. Nine fields
cover the cluster core, the tenth is positioned to obtain high resolution data on NGC 1404 and
to double the exposure in the region between NGC 1404 and NGC 1399.\label{fig1}}
\end{figure}

\begin{figure}
%\plotone{f2_compress.eps}
\caption{Soft band (0.3-1.5 keV), adaptively smoothed, exposure corrected image of Fornax mosaic. \label{fig2}}
\end{figure}

\begin{figure}
%\plotone{f3_compress.eps}
\caption{Medium band (1.5-2.5 keV) image of Fornax mosaic.\label{fig3}}
\end{figure}

\begin{figure}
%\plotone{f4_compress.eps}
\caption{Hard band (2.5-8.0 keV) image of Fornax mosaic.\label{fig4}}
\end{figure}

\begin{figure}
%\plotone{f5_compress.eps}

\caption{Gas temperature map of Fornax, made using the algorithm
described in the text. The majority of the gas emission is cooler than
$\sim 1.6$ keV in agreement with previous measurements. Arrows
indicate the suggested direction of motion of the cluster member
galaxies NGC 1404 (East) and NGC 1387 (West) from the apparent cool
``tails'' of these objects. }

\end{figure}

\begin{figure}
%\plotone{f6_compress.eps}
\caption{Left panel: (a) Contours of net photon counts,
uncorrected for exposure, are overlaid on the temperature map. Spacing is logarithmic,
lowermost contour corresponds to the enclosure of pixels with $\sim
60$ counts. Temperatures outside these regions should be treated as
poorly constrained. Right panel: (b) Narrow band (140 eV width) X-ray image centered on the Fe-L complex, with
continuum subtracted.}
\end{figure}

\begin{figure}
%\plotone{f7_compress.eps}
\caption{Pie-wedge annuli used to extract the NW radial emission profile of NGC 1404, overlaid on
soft band image. Irregular shaped region at edge of annuli is the region used to estimate the
local background, which avoids point sources and chip gaps, but includes emission due to the Fornax
core region. \label{fig5}}
\end{figure}

\begin{figure}
\plotone{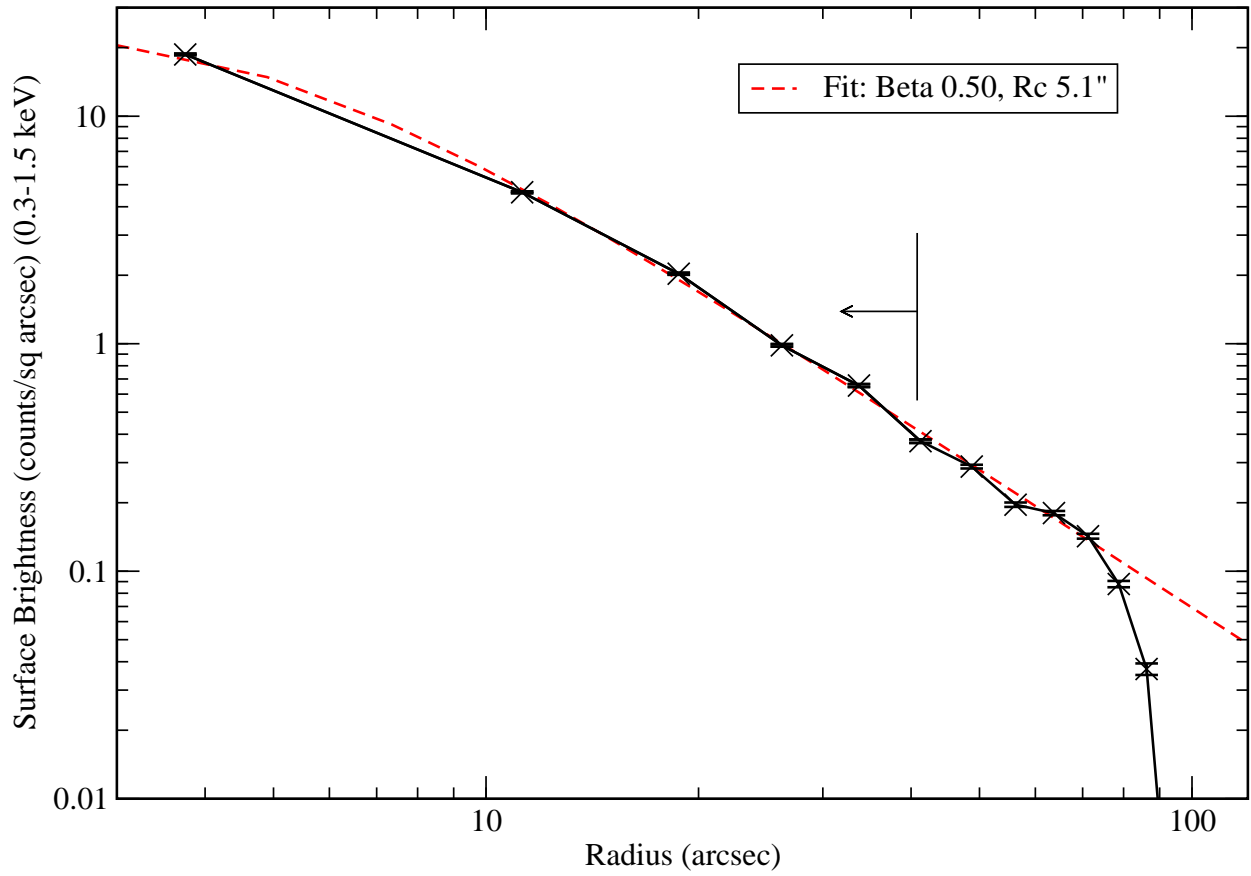}
\caption{Surface brightness profile of NW quadrant of NGC 1404 in 0.3-1.5 keV band. Background has been subtracted. \label{fig6}}
\end{figure}

\begin{figure}
\plotone{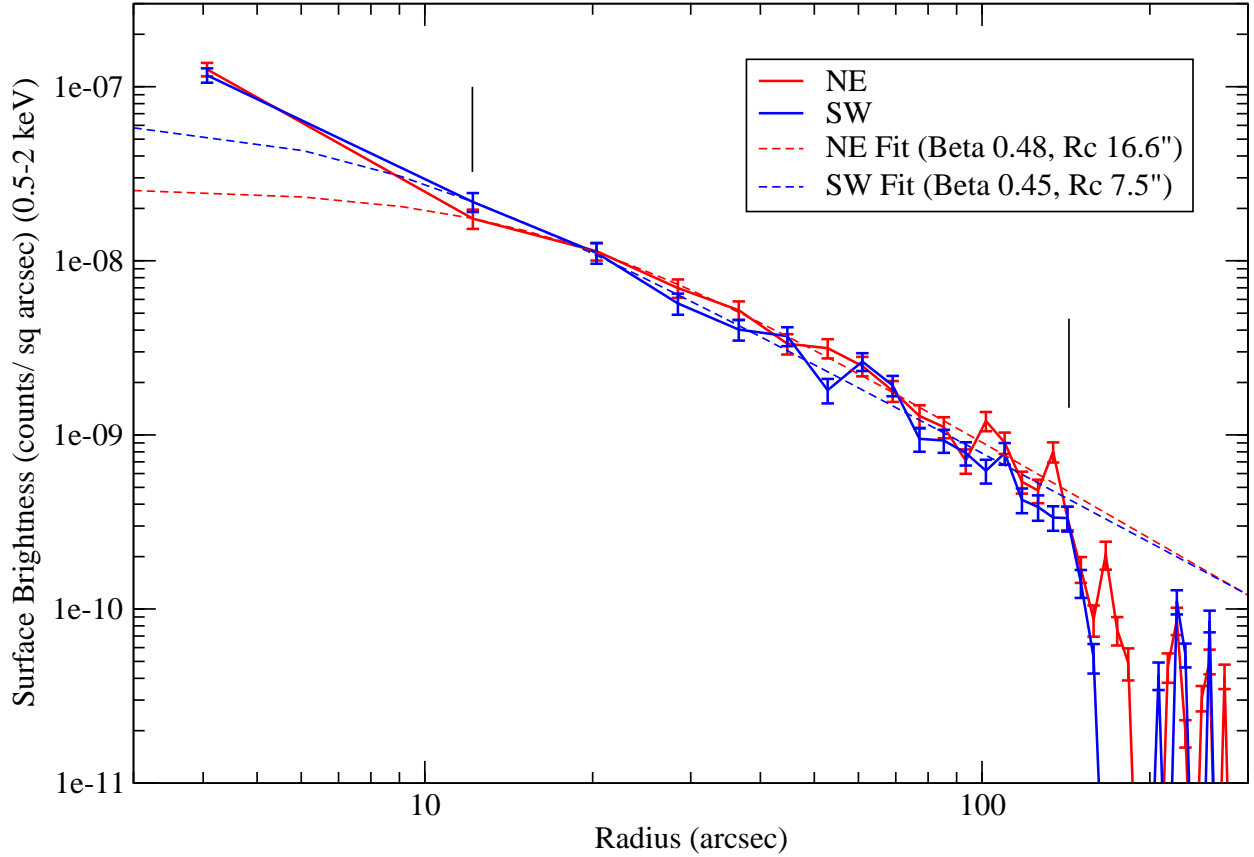}
\caption{Surface brightness profile of NE and SW regions of NGC 1387 in 0.5-2 keV band (to maximise photon count). Similar annuli
to those used for NGC 1404 (Figure 7) are used here, orientated to the NE and SW respectively.
}            
\end{figure}

\begin{figure}
%\plottwo{f10a_compress.eps}{f10b_compress.eps}
\caption{Left panel: All source detections (blue ellipses) overlaid on smoothed soft band image, color scale is logarithmic
and has been stretched to illustrate overall morphology of the ICM. Right panel: $7'\times 7'$ close-up 
of un-smoothed broad band (0.3- 8 keV) data (1'' pixels) centered on NGC 1399 with source detections overlaid.  }            
\end{figure}

\begin{figure}
%\plotone{f11.eps}
\caption{X-ray color-color plot of 437 sources with CFS ACIS-I flux in
each of 3 bands; soft (0.3-1.0 keV), medium (1.0-2.0 keV), hard (2.0-8.0 keV).
Plotted curves correspond to varying source spectral types. Curves are functions of
photo-electric absorption, lowermost points on curves correspond to the
known Galactic foreground absorption of $nH=1.4\times 10^{20}$ cm$^{-2}$, increasing to
a maximum of $nH=2\times 10^{22}$cm$^{-2}$.}            
\end{figure}

\begin{figure}
%\plotone{f12_compress.eps}
\caption{The positions of all X-ray sources and bright cluster galaxy members overlaid on a
soft band image of Fornax. Arrows indicate galaxies which are within 3 kpc (30'') of an X-ray
source.}            
\end{figure}

\begin{figure}
%\plotone{f13_compress.eps}
\caption{Overview of the larger environment of Fornax. Approximately 1 Mpc to the South-West of the
Fornax core (shown by the scaled soft-band {\sl Chandra} image overlaid on a wide-field optical image) is a group of at least 16 galaxies identified by ~\citet{dri01} using the 2dF spectroscopic Fornax survey data as infalling substructure. The putative infall direction is labeled by an arrow. The axis between this group and the center of Fornax defined by NGV 1399 also appears to correspond to an axis defined by the ``swept'' back plume of ICM to the North-East of NGC 1399.}
\end{figure}

\begin{figure}
%\plotone{f14_compress.eps}
\caption{Locations of the two bright GC candidates with X-ray detections from ~\citet{mie02} are
shown relative to the soft band image of Fornax. (A) - FCOS 1-060, (B) - FCOS 2-073.}            
\end{figure}

\begin{figure}
%\plotone{f15_compress.eps}
\caption{Left panel: contours of the adaptively smoothed 0.3-1.5 keV soft band image of the possible background galaxy cluster
to the East of NGC 1399 are
overlaid on the raw soft band data, binned to 4'' pixels. Right panel: X-ray contours overlaid on
the deep R-band CTIO 4-m image (D. Neill). The heavy circle indicates the change in scale between these
two panels and is centered on the potential cluster in both cases.}            
\end{figure}

\clearpage

\begin{deluxetable}{lcccccc}
\tabletypesize{\scriptsize}
\tablecaption{X-ray properties of detected globular cluster candidates in outer NGC 1399 field}
\tablewidth{0pt}
\tablehead{
\colhead{GC} & \colhead{Counts}   & \colhead{ct/s}   &
\colhead{Soft index} &
\colhead{Hard index}  & \colhead{Flux (erg s$^{-1}$ cm$^{-2}$)} & \colhead{Luminosity (erg s$^{-1}$)}
}
\startdata
FCOS 1-060 & $60\pm 11$ & $7\pm 1 \times 10^{-4}$ & -0.35 & 0.39 & $4\pm 0.6 \times 10^{-15}$ & $2\pm 0.3\times 10^{38}$\\
FCOS 2-073 & $17\pm 4$ & $4\pm 1 \times 10^{-4}$ & -0.1 & 0.24 & $4 \pm 1.0 \times 10^{-15}$ & $2\pm 0.5\times 10^{38}$\\
 \enddata

\tablecomments{All counts, rates, fluxes and luminosities are quoted in the 0.3-8 keV
band. Galactic absorption is assumed, $nH=1.4 \times 10^{20}$ cm$^{-2}$. Nomenclature from
Table 2, ~\citet{mie02}}

\end{deluxetable}

\clearpage 

%% If you are not including electonic art with your submission, you may
%% mark up your captions using the \figcaption command. See the 
%% No more than seven \figcaption commands are allowed per page, 
%% so if you have more than seven captions, insert a \clearpage 

%% Tables should be submitted one per page, so put a \clearpage before

%% Two options are available to the author for producing tables:  the
%% deluxetable environment provided by the AASTeX package or the LaTeX
%% table environment.  Use of deluxetable is preferred.
%%

%% Three table samples follow, two marked up in the deluxetable environment,
%% one marked up as a LaTeX table.

%% In this first example, note that the \tabletypesize{}
%% command has been used to reduce the font size of the table.
%% Note also that the \label command needs to be placed 

%% Tables may also be prepared as separate files. See the accompanying
%% sample file table.tex for an example of an external table file.
%% To include an external file in your main document, use the \input
%% command. Uncomment the line below to include table.tex in this
%% sample file. (Note that you will need to comment out the \documentclass,
%% \begin{document}, and \end{document} commands from table.tex if you want
%% to include it in this document.)


%% The following command ends your manuscript. LaTeX will ignore any text

\end{document}